\documentclass[twocolumn,prl,nobibnotes,superscriptaddress,floatfix]{revtex4-2} 

\usepackage[dvipdfmx]{graphicx}
\usepackage{amsmath,amssymb,bm}
\usepackage{xcolor}
\usepackage{float}
\usepackage{ulem}
\usepackage{lipsum, babel}
\usepackage[
colorlinks=true,
citecolor=blue,
setpagesize=false]{hyperref}

\newcommand{\cyan}[1]{{\color{cyan}{\it {#1}}}} 

\begin{document}

\title{
Collective non-Hermitian skin effect: Point-gap topology and the doublon-holon excitations in non-reciprocal many-body systems
}

\author {Beom Hyun Kim}
\affiliation{Center for Theoretical Physics of Complex Systems, Institute for Basic Science, Daejeon 34126, Republic of Korea}

\author {Jae-Ho Han}
\affiliation{Center for Theoretical Physics of Complex Systems, Institute for Basic Science, Daejeon 34126, Republic of Korea}

\author {Moon Jip Park}
\email{moonjippark@hanyang.ac.kr}
\affiliation{Department of Physics, Hanyang University, Seoul 04763, Republic of Korea}

\date{\today} 

\begin{abstract}
Open quantum systems provide a plethora of exotic topological phases of matter that has no Hermitian counterpart. Non-Hermitian skin effect, macroscopic collapse of bulk states to the boundary, has been extensively studied in various experimental platforms. However, it remains an open question whether such topological phases persist in the presence of many-body interactions. Notably, previous studies have shown that the Pauli exclusion principle suppresses the skin effect. In this study, we present a compelling counterexample by demonstrating the presence of the skin effect in doublon-holon excitations. While the ground state of the spin-half Hatano-Nelson model shows no skin effect, the doublon-holon pairs, as its collective excitations, display the many-body skin effect even in strong coupling limit. We rigorously establish the robustness of this effect by revealing a bulk-boundary correspondence mediated by the point gap topology within the many-body energy spectrum. Our findings underscore the existence of non-Hermitian topological phases in collective excitations of many-body interacting systems.
\end{abstract}

\maketitle

\cyan{Introduction--} Dissipation is a ubiquitously observed phenomenon in many physical systems. In quantum systems, the presence of the dissipation leads to complex-valued eigenenergies, which mark the onset of non-Hermitian quantum mechanics. Non-reciprocal interactions, which are typical in out-of-equilibrium phenomena, are another important source of non-Hermiticity. The broad applicability of non-Hermitian quantum mechanics has attracted considerable interest in various fields of physics including physics, such as high-energy~\cite{Bender2007,Jones2010,Bender2005,Alexandre2015}, optics~\cite{Feng2017,Ozdemir2019,Miri2019,Ozawa2019,Longhi2020,Zhu2020}, cold atoms~\cite{Takasu2020,Guo2022,Liang2022,Zhou2022,Hamanaka2023}, and condensed matter systems~\cite{Hatano1996,Hatano1998,Fukui1998,Kozii2017,Yoshida2018,Nagai2020,Rausch2021,Ren2022}.

Recent progress in the field of non-Hermitian topological phases has shown great promise in discovering new types of topological phases~\cite{Gong2018,Yao2018,Yao2018b,Kawabata2019,Song2019,Yokomizo2019,Deng2019,Okuma2020,Borgnia2020,Bergholtz2021,Ding2022,Lin2023,Kawabata2023}. The non-Hermitian skin effect (NHSE), which exhibits the macroscopic collapse of the eigenstates to the boundary, is the representative example~\cite{Lee2019,Jiang2019,Imura2019,Zhang2020,Yang2020,Yi2020,Schomerus2020,Ghaemi2021,Ghatak2020,Helbig2020,Weidemann2020,Jahan2021,Longhi2021,Kim2021,Gu2022,Wang2022}. However, the fate of the NHSE in the presence of the many-body interaction is still elusive. Previous studies reveal that the NHSE can be fragile against many-body effects~\cite{Lee2020,Zhang2022,Kawabata2022,Alsallom2022,Liu2020,Longhi2023}. For instance, in the half-filled interacting Hatano-Nelson model, the macroscopic accumulation of charge at the boundary is generally prohibited by the Pauli exclusion principle.

In this work, we firstly establish the case that the NHSE robustly exists as a form of collective excitations. Unlike spinless models, the antiferromagnetic Mott insulator ground states of the spin-half Hatano-Nelson model harbor the doublon-holon pairs as well-defined collective excitations even in strong interacting limits. While the ground state does not exhibit the NHSE, we show that the excited states with the doublon-holon pairs show the helical skin effect, where the doublon and the holon exhibit the counterpropagating accumulations of the charge at the opposite boundaries. We formally establish this exotic collective NHSE by illustrating bulk-boundary correspondence mediated by point-gap topology in the many-body energy spectra.

\cyan{Lattice Model and Symmetries--}
We consider the one-dimensional interacting Hatano-Nelson model 
of spin-half fermions, which is described by the following Hamiltonian:
\begin{align}
H =& - t \sum_{l=0}^{L-2}\sum_{\sigma}
\left(
e^{A} c^{\dagger}_{l+1,\sigma}  c_{l,\sigma} +
e^{-A} c^{\dagger}_{l,\sigma}  c_{l+1,\sigma} 
\right) \nonumber \\
& + U \sum_{l=0}^{L-1} n_{l,\uparrow} n_{l,\downarrow} + H_{B},
\label{Eq_Hubbard}
\end{align}
Here, $c^{\dagger}_{l,\sigma}$ ($c_{l,\sigma}$) are the fermion creation (annihilation) operators with spin $\sigma$ ($=\uparrow,\downarrow$) at the $l$-th site, $n_{l,\sigma}=c^{\dagger}_{l,\sigma}c_{l,\sigma}$ is the number operator, and $U$ is the strength of on-site Coulomb repulsion strength between doubly occupied fermions. The lattice size is denoted by $L$ and $A$ represents the imaginary vector potential that introduces non-reciprocal hopping~\cite{Hatano1996,Hatano1998}. The boundary term $H_B$ accounts for the hopping Hamiltonian between the end sites. For example,
in open boundary conditions (OBC), $H_{B}$ is $0$,
while $H_{B}=-t \left( e^{A} c^{\dagger}_{0,\sigma} c_{L-1,\sigma} +  e^{-A} c^{\dagger}_{L-1,\sigma}c_{0,\sigma} \right)$ for periodic boundary conditions (PBC). In general twisted boundary conditions (TBC), $H_{B}$ takes a form, $-t \left( e^{A+i\phi} c^{\dagger}_{0,\sigma} c_{L-1,\sigma} + e^{-A-i\phi} c^{\dagger}_{L-1,\sigma} c_{0,\sigma} \right)$, where $\phi$ represents the $U(1)$-gauge (magnetic) flux.

The symmetry of the Hamiltonian depends on both boundary conditions and the presence of the many-body interactions. In the non-interacting limit ($U=0$), the Hamiltonian under PBC and TBC satisfies the normality ($HH^\dagger =H^\dagger H $), which allows to have orthonormalized eigenstates. The corresponding eigenvalue spectra in PBC can form a closed contour with non-trivial point gap topology. In the corresponding OBC, the normality is lost. Instead, there exists the similarity transformation with the invertible operation $S$ 
that satisfies $H_{A=0} = S^{-1} H S$, where $H_{A=0}$ is the Hermitian Hamiltonian without the imaginary gauge flux $A$ (see Supplement Materials (SM)~\cite{Suppl}). As a result, all eigenvalues exhibit the purely real numbers~\cite{Mostafazadeh2002b}. The collapse of the eigenenergies to the real axis manifests as the NHSE.

Even in the finite many-body interaction ($U\neq 0$), we can find the similarity transformation $S$ such that $H_{A=0} = S^{-1} H S$ under OBC, resulting the purly real eigenvalues. However, the nomality is broken regardless of the boundary condition due to the incusion of $U$. Morever, $H$ for PBC and TBC becomes $\mathcal{PT}$-pseudo-Hermitian, where $\mathcal{P}$ and $\mathcal{T}$ are inversion and time-reversal operators, respectively~\cite{Suppl}. Consequently, the eigenvalues are either real or occur in complex-conjugate pairs~\cite{Mostafazadeh2002a}.

\begin{figure}[b]
\centering
\includegraphics[width=0.9\columnwidth]{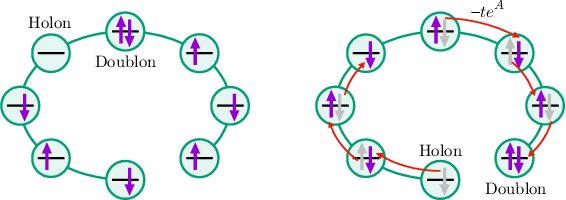}
\caption {
Schematic diagrams of the motion of the doublon (doubly occupied site) and holon (empty site) in the present of non-reciprocal hopping. As illustrated on the right panel, non-reciprocal hopping, denoted by red arrows, results in the rightward (leftward) movement of doublons (holons). The gray arrows represent the initial positions of fermion spins prior to hopping. Under OBC, doublon and holon become localized at opposite edges of one dimensional chain. This segregation in localized doublon and holon is responsible for the NHSE.
}
\label{fig_propDH}
\end{figure}

\cyan{Non-Hermitian skin effect--} We represent the many-body eigenstates using orthonormal basis states: $\left| \Psi_n \right> = \prod_{j=1}^N
c^{\dagger}_{l_{nj},\sigma_{nj}}\left| vac \right>$,
where $l_{nj}$ and $\sigma_{nj}$ denote the site index and spin state of
the $j$th fermion for the $n$th basis state, respectively.
$\left| vac \right>$ represents the vacuum state,
and $N$ is the total number of fermions. 
The similarity transformation $S$ such that $H_{A=0} = S^{-1} H S$ under OBC is given as following~\cite{Suppl}:
\begin{equation}
S = \sum_n e^{A \sum_{j=1}^N l_{nj}} \left| \Psi_n \right>\left< \Psi_n \right|.
\end{equation}
The non-orthonormal factor $C_n \left(A\right) = e^{A \sum_{j=1}^N l_{nj}}$ causes coefficients of eigenstates to vary exponentially with $A$, playing a crucial role in determining the NHSE in many-body systems~\cite{Alsallom2022,Heussen2021,Suppl}.


\begin{figure*}[!t]
\centering
\includegraphics[width=1.8\columnwidth]{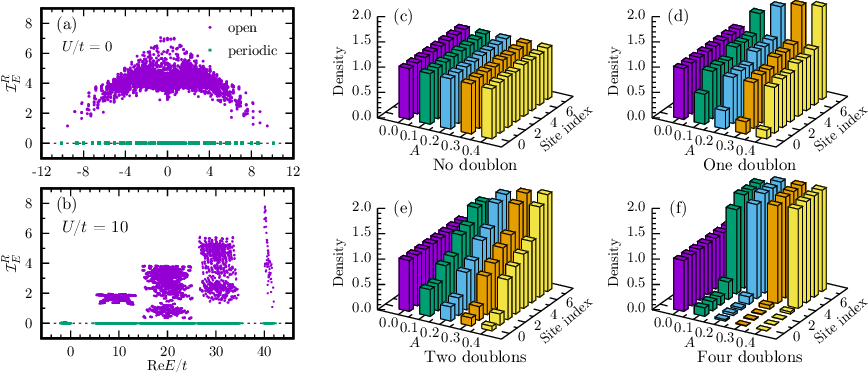}
\caption {
The number density imbalance in the half-filled Hatano-Nelson model of spin-half fermions ($A=0.3$) under both OBC and PBC when (a) $U/t=0$ and (b) $U/t=10$.
The imbalance $\mathcal{I}_E^R$ is calculated by $\sum_{L/2 \le l < L} n_E^R \left( l \right) - \sum_{0 \le l < L/2} n_E^R \left( l \right)$, where
$n_E^R \left( l \right)$ is the number density of right eigenvalues at the $l$th site.
(c)--(f) the number density distributions as a function of imaginary gauge potential $A$ for specific right eigenstates, where no, one, two, and four doublon-holon pairs play a dominant role for OBC. All results are obtained for a lattice size of eight ($L = 8$).
}
\label{fig_skin}
\end{figure*}

With respect to the half-filled ground state ($N=L$), $C_n(A)$ can vary depending on the distribution of doubly occupied sites, called doublons, and empty sites, referred to as holons~\cite{Suppl}. Explicitly,
let $N_{d}$ be the total number of doublon-holon pairs.
$C_n \left(A\right)$ is expressed as
\begin{equation}
C_n \left(A\right) = e^{A\frac{L(L-1)}{2}}
e^{A \sum_{j=1}^{N_{d}} \left( l^d_{nj} - l^h_{nj} \right)},
\end{equation}
where $l^d_{nj}$ and $l^h_{nj}$ denote the site indices of
the $j$th doublon and holon for the $n$th basis state $\left| \Psi_n \right>$, respectively.
For states with no doublon-holon pair ($N_d=0$),
$C_n \left(A\right)$ reaches its minimum with a constant value of $e^{A\frac{L(L-1)}{2}}$.
These states exhibit little influence from the NHSE.
On the other hand, for states with finite doublon-holon pairs,
$C_n \left(A\right)$ grows exponentially as the strength of $A$ increases.
The larger the segregation between doublons and holons,
quantified by $\sum_{j=1}^{N_{d}} \left( l^d_{nj} - l^h_{nj} \right)$,
the larger $C_n \left(A\right)$ becomes.
Hence, the NHSE in the half-filled case is characterized by the segregation of doublons and holes.

The relation between the NHSE and the segregation of doublons and holons can be explained by considering the non-reciprocal mobility of these particles.
As illustrated in Fig.~\ref{fig_propDH}, let us imagine that a doublon-holon pair forms between neighboring sites in a half-filled state.
When a positive $A>0$ is present, the hopping strength from the left site to the right site becomes larger than the opposite direction. This enhances the fermion on the left side of the holon site to preferentially occupy that site, leading to the holon's movement to the right.
In contrast, one of the fermions occupying the doublon site is more likely to hop to the next site on the right, resulting in the doublon's movement to the left.
As a result, doublons and holons become separated and localized at opposite boundary sites under OBC. Conversely, for PBC, doublons and holons merge again.
This distinctive non-reciprocal behavior of doublons and holons underpins the emergence of the NHSE in the half-filled case.

In many-body systems governed by the Pauli exclusion principle, one effective approach to demonstrate the NHSE is by quantifying the asymmetry in the distribution of number density along an open-boundary chain~\cite{Alsallom2022}.
To achieve this, we calculate the number density distribution of the right eigenstates using the formula: $n_E^R \left( l \right) = 
\left< \Psi_E^R \right|\sum_{\sigma} c^\dagger_{l,\sigma} c_{l,\sigma} \left| \Psi_E^R \right>/\left< \Psi_E^R \big| \Psi_E^R \right>$.
This distribution's asymmetry results in the number imbalance of fermions located at below $L/2$ ($0 \le l < L/2$) and above $L/2$ ($L/2 \le l < L$) sites. We can quantify this imbalance with the following formula:
\begin{equation}
\mathcal{I}_E^R =  \sum_{L/2 \le l < L} n_E^R \left( l \right) -
\sum_{0 \le l < L/2} n_E^R \left( l \right).
\end{equation}
The occurrence of non-zero values for $\mathcal{I}_E^R$ is a distinctive indicator of the many-body NHSE, as detected within right eigenstates.

Fig.~\ref{fig_skin} exhibits the calculated number density distribution, clearly showing that the asymmetric nature is present under OBC for both non-interacting ($U = 0$) and interacting ($U \ne 0$) cases. The corresponding finite values of imbalance $\mathcal{I}_E^R$ affirm this. However, the situation differs for PBC, where $\mathcal{I}_E^R$ consistently equals zero. These results depict the NHSE in non-reciprocal many-body systems.

In the non-interacting case ($U=0$), $\mathcal{I}_E^R$ exhibits an arch-like pattern, ranging from about $1$ to roughly $7$ for $L=8$. In contrast, strong interactions ($U=10t$) yield a stair-like $\mathcal{I}_E^R$ pattern. This can be understood by the spectral properties of the Mott insulating region. With a substantial Coulomb repulsion ($U$), eigenvalues mainly depend on $U$ strength, leading to sizable separations related to the number of doublon-holon pairs ($N_d$). When $N_d$-dominated states prevail in the eigenstates, $\mathcal{I}_E^R$ attains a maximum value of around $2N_d$ for OBC. This feature is well captured by Fig.~\ref{fig_skin}(b).

Furthermore, the non-reciprocal hopping causes opposing doublon and holon propagation. Initially, they gather at opposing boundaries for finite $A$. The Pauli exclusion principle then drives further accumulation near by initial sites. This pattern is visually evident in the calculated number density distributions, depicted in Figs.~\ref{fig_skin}(c)--(f). Consequently, we deduce that the NHSE in the many-body system can be understood through the segregation of doublons and holons within the half-filled Hatano-Nelson model of spin-half fermions.

\begin{figure}[!t]
\centering
\includegraphics[width=0.9\columnwidth]{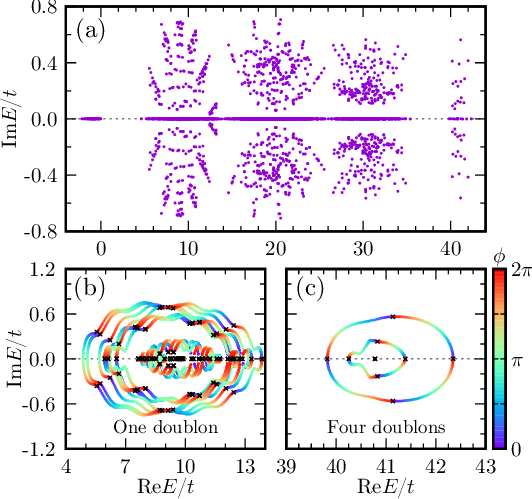}
\caption {
 (a) Eigenvalue spectrum in the half-filled Hatano-Nelson model of spin-half fermions with $U/t=10$ and $A=0.3$. Notably, even under strong interaction limit, complex eigenvalues remain robust within finite doublon-holon sectors.
Spectral evolution of eigenvalues as functions of the gauge flux $\phi$ for TBC in (b) one and (c) four doublon-holon sectors. For simplicity, only eigenvalues corresponding to the crystal momentum $k=0$ and the $z$-component of the total spin $S_z=0$ are presented in (b) and (c) by utilizing the translation symmetry and $S_z$ conservation (see SM~\cite{Suppl}). Cross points indicate eigenvalues at $\phi=0$. These spectral data effectively depict the point-gap topology of complex eigenvalues. All results are obtained for a lattice size of eight ($L = 8$).
}
\label{fig_energy}
\end{figure}

\cyan{Complex eigenspectrum and point-gap topology--}
As shown in Fig.~\ref{fig_energy}(a), for PBC, the eigenvalues of the Hamiltonian can take either on purely real values or form complex conjugate pairs. This originates from the pseudo-Hermiticity of the Hamiltonian. To gain a deeper understanding of the spectral properties, we analyze how eigenvalues change with the gauge flux $\phi$ under TBC. Figs.~\ref{fig_energy}(b) and (c) provide the trajectory of complex eigenvalues when the real part of the eigenvalue lies between $4t$ and $14t$ (one doublon-holon sector) and between $39t$ and $43t$ (four doublon-holon sector), respectively.

As $\phi$ changes from $0$ to $2\pi$, complex eigenvalues undergo rotations in the complex plane, effectively tracing closed paths alongside other eigenvalues. This rotation ultimately results in non-zero integer winding numbers, which can be calculated using the formula:
\begin{equation}
W \left(E_p\right) = \oint_0^{2\pi} \frac{d \phi}{2\pi i} \frac{d}{d \phi}
\log \det \left[ H(\phi) - E_p \right],
\label{Eq_WN}
\end{equation}
Here, $E_p$ stands for the point gap of the complex energy~\cite{Zhang2022,Kawabata2022}. This behavior highlights the intriguing topological aspects inherent to the system.

In the non-interacting case, all complex eigenvalues wind around the origin ($E=0$). It implies that the point gap is straightforwardly identified at the center of the complex plane (see SM~\cite{Suppl}).
However, in the interacting case, as seen in Fig. \ref{fig_energy}(b) and (c), multiple point gaps are not isolated but emerge near the energy center of doublon-holon sectors. This multiplicity of point gaps defines the non-Hermitian topology of non-reciprocal many-body systems.

In the strong interaction limit, the many-body NHSE does not emerge for eigenstates mainly contributed to states with no doublon-holon pair in the half-filled case.
Correspondently, eigenvalues below zero remain purely real even for PBC.
In the second-order perturbation limit, 
the effective Hamiltonian in no doublon-holon sector is described with
the Heisenberg interaction, given as $H_{\rm eff} = J \sum_{l} \mathbf{S}_l
\cdot \mathbf{S}_{l+1}$~\cite{Essler2005}.
The magnetic interaction among neighboring spins ($J$) is determined by virtual hopping processes, in which fermions hop forth and back neighboring sites.
$J$ is expected to be robust as $J = \frac{4t^2 e^{A} e^{-A}}{U} =\frac{4t^2}{U}$.
Therefore, the real eigenvalues are stabilized for no doublon-holon sector
in the strong interaction limit (see SM~\cite{Suppl}).

When the total number of doublon-holon pairs are $L/2$,
$L/2$ number of doublons and holons occupy all lattice sites.
In this situation, we can construct the effective model using 
attracted half-filled hard-core bosons with non-reciprocal hopping
in the second-order perturbation limit.
The effective attracted interaction, hopping strength, and 
effective imaginary gauge are estimated by
$U_{\rm eff}=-\frac{4t^2}{U}$, $t_{\rm eff}=\frac{2t^2}{U}$, and $A_{\rm eff} = 2A$, respectively~(see SM~\cite{Suppl}).
Since $\left| U_{\rm eff} \right| / t_{\rm eff} = 2$ is much smaller than $U/t$ and $A_{\rm eff}$ is the twice of $A$, the robustness of complex eigenvalues can be maintained even when $U$ exceeds $t$ much largely.

\begin{figure}[!t]
\centering
\includegraphics[width=0.9\columnwidth]{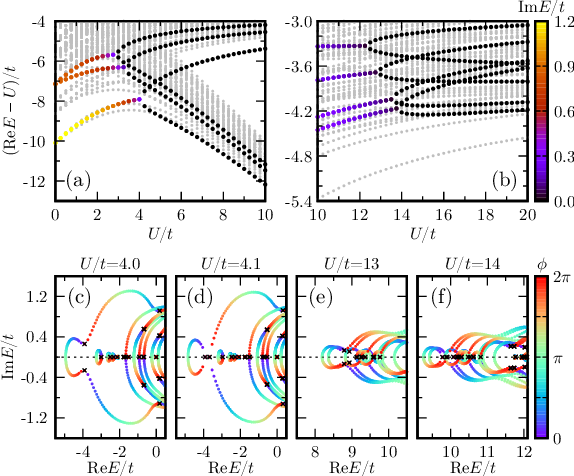}
\caption {
Real components of eigenvalues varying with Coulomb repulsion $U$ for (a) $0\le U/t \le 10$ and (b) $10\le U/t \le 20$.
(a) and (b) represent the eigenvalue spectrum for no and one doublon-holon sector, respectively. We highlight a few eigenpairs, which perform the transition from complex conjugate pair to two pure real numbers, using color map.
(c)--(f) show the spectral evolutions of eigenvalues of $k=0$ and $S_z=0$ eigenstates as a function of the gauge flux $\phi$ under TBC. All results are obtained when $A=0.3$ and $L=8$.
}
\label{fig_trans}
\end{figure}

\cyan{Topological transition and exceptional points--}
Due to the Hermitian nature of the Coulomb interaction part in Eq.~\ref{Eq_Hubbard}, all eigenvalues asymptotically converge to real numbers in the $U$-infinite limit even for PBC.
All complex conjugate eigenpairs in finite $U$ eventually transform into two real eigenvalues.
This transition involves the gradual separation of two real eigenvalues, following the merging of complex conjugate eigenpairs along the real axis. 
This spectral behavior resembles the $\mathcal{PT}$-phase transition~\cite{Longhi2019,Longhi2019b}. Figure~\ref{fig_trans}(a) and (b) well depict the sequential occurrences of these transitions when $U$ gradually increases.

We further examine the evolution of complex eigenvalues in the presence of the gauge flux $\phi$ under TBC.
As shown in Fig.~\ref{fig_trans}(c)--(f), complex conjugate eigenpair are connected by the $\phi$ evolution before the transition,
while that eigenpair get separated after the transition.
It implies the winding number in Eq.~\ref{Eq_WN} changes undergoing the transition,
giving the different point-gap topology~\cite{Longhi2019,Alsallom2022}.
Moreover, two eigenstates of the eigenpair involved in the transition collapse into a singular state at the transition point (see SM~\cite{Suppl}). 
The transition point of the winding number is identified as $\mathcal{PT}$-like topological transitions accompanying the exceptional point. (see SM~\cite{Suppl} for the topological transitions as a function of $A$).

\cyan{Summary and discussion--}
We investigate the intriguing behavior of the many-body NHSE and unique features of eigenspectra in the half-filled interaction Hatano-Nelson model of spin-half fermions. We discover that strong interactions suppress doublon-holon excitaitons in the ground state, leading to the absence of the NHSE. However, excited eigenstates still exhibit these excitations, driven by non-reciprocal hopping, which causes doublons and holons to move in opposite directions. The spatial segregation of doublon-holon pairs serves the hallmark of the many-body skin modes. Moreover, these modes are attributed to the bulk-boundary correspondence mediated by the point gap topology within complex many-body eigenspectrums.

Experimentally, realizing these effects can be challenging, but open quantum systems such as ultracold atoms can be a promising platform. Experiments with bosonic ultracold atoms have already shown skin effects~\cite{Liang2022}. Because the dynamics of the systems is governed by the Liouvillian master equations, exploring the interplay between many-body effects and Liouvillian skin effects~\cite{Hamanaka2023,Haga2021,Yang2020,Li2023} in such systems holds great potential for unveiling novel physical phenomena of non-Hermitan many-body physics.

\begin{acknowledgments}
\cyan{Acknowledgements--}
We acknowledge Jung-Wan Ryu, Hee Chul Park, and Sonu Verma for fruitful discussion. This work was supported from the Institute for Basic Science in the Republic of Korea through the Project No. IBS-R024-D1. This work was supported by the National Research Foundation of Korea
(NRF) grant funded by the Korea government (MSIT) (Grants No. RS-2023-00252085 and No. RS-2023-00218998).
\end{acknowledgments}

\bibliography{references}

\newpage

\setcounter{table}{0}
\setcounter{figure}{0}
\setcounter{equation}{0}

\begin{widetext}
\begin{center}
	{\bf \Large
		\textit{Supplemental Material}:\\
		Point-gap topology and skin effect of the doublon-holon excitations in non-reciprocal many-body systems
	}
	
	\vspace{0.2 cm}
\end{center}

\section{Hamiltonian properties}

\subsection{Hamiltonian}

We consider the one-dimensional  Hatano-Nelson model of spin-half fermions
described by the following Hamiltonian:
\begin{equation}
	H = - t \sum_{l=0}^{L-2}\sum_{\sigma}
	\left(
	e^{A} c^{\dagger}_{l+1,\sigma}  c_{l,\sigma} +
	e^{-A} c^{\dagger}_{l,\sigma}  c_{l+1,\sigma} 
	\right)
	+ U \sum_{l=0}^{L-1} n_{l,\uparrow} n_{l,\downarrow} + H_B 
	\label{Eq_Hubbard_all}
\end{equation}
where $c^{\dagger}_{l,\sigma}$ ($c_{l,\sigma}$) represents 
the creation (annihilation) operator of a fermion 
with spin $\sigma$ ($=\uparrow,\downarrow$) at the $l$th site,
and $n_{l,\sigma}=c^{\dagger}_{l,\sigma}c_{l,\sigma}$ is the number operator at the site.
The lattice size is denoted by $L$ and $A$ refers to the imaginary vector potential 
to give rise to the non-reciprocal hopping~\cite{Hatano1996,Hatano1998}.
Additionally, $H_B$ accounts for the hopping Hamiltonian between fermions at the boundary and can be adopted differently for periodic, open, and
twisted boundary conditions as follows:
$H_{B}$ can are given as
\begin{subequations}
	\begin{equation}
		H_{B}^{PBC} = -t \sum_\sigma 
		\left( e^{A} c^{\dagger}_{0,\sigma}  c_{L-1,\sigma} +
		e^{-A} c^{\dagger}_{L-1,\sigma}c_{0,\sigma} \right),
	\end{equation}
	\begin{equation}
		H_{B}^{OBC} = 0,
	\end{equation}
	\begin{equation}
		H_{B}^{TBC} = -t \sum_\sigma
		\left( e^{A+i\phi} c^{\dagger}_{0,\sigma}  c_{L-1,\sigma} +
		e^{-A-i\phi} c^{\dagger}_{L-1,\sigma}  c_{0,\sigma} \right),
	\end{equation}
\end{subequations}
where $\phi$ denotes the $U(1)$-gauge (magnetic) flux.
For twisted boundary condition, we can perform 
a unitary transformation of local fermionic operators 
as follows:
\begin{equation}
	c^{\dagger}_{l\sigma} \rightarrow
	\tilde{c}^{\dagger}_{l\sigma} = e^{i l\phi/L} c^{\dagger}_{l\sigma}, \quad
	c_{l\sigma} \rightarrow \tilde{c}_{l\sigma} = e^{-i l\phi/L} c_{l\sigma}.
\end{equation}
This transformation allows us to obtain an effective periodic Hamiltonian as follows:
\begin{equation}
	H=- t \sum_{l=0}^{L-2}\sum_{\sigma}
	\left(
	e^{A+i\phi/L} \tilde{c}^{\dagger}_{l+1,\sigma}  \tilde{c}_{l,\sigma} +
	e^{-A-i\phi/L} \tilde{c}^{\dagger}_{l,\sigma}  \tilde{c}_{l+1,\sigma} 
	\right) 
	+ U \sum_{l=0}^{L-1} \tilde{n}_{l,\uparrow} \tilde{n}_{l,\downarrow}
	+ \tilde{H}_{B},
\end{equation}
where $\tilde{n}_{l,\sigma}= 
\tilde{c}^{\dagger}_{l,\sigma}\tilde{c}_{l,\sigma}$ and
$\tilde{H}_{B}
= -t \sum_\sigma \left( e^{A+i\phi/L} 
\tilde{c}^{\dagger}_{0,\sigma}  \tilde{c}_{L-1,\sigma} +
e^{-A-i\phi/L} \tilde{c}^{\dagger}_{L-1,\sigma} \tilde{c}_{0,\sigma} \right)$.

\subsection{Pseudo-Hermiticity}

Similar to the spinless Hatano-Nelson model~\cite{Zhang2022}, 
the Hamiltonian of Eq.~\ref{Eq_Hubbard_all} can exhibit 
the pseudo-Hermiticity for the twisted (periodic) boundary condition
with respect to the $\mathcal{PT}$ symmetry.
Let $\mathcal{P}$ and $\mathcal{T}$ be inversion and time-reversal operators, respectively.
We can get the following relations:
\begin{equation}
	\left( \mathcal{PT} \right) c_{l,\sigma}^\dagger 
	\left( \mathcal{PT} \right)^{-1} = 
	\left( -1 \right)^{n_\sigma} c_{L-1-l,\bar{\sigma}}^\dagger, \quad
	\left( \mathcal{PT} \right) e^{i\phi}
	\left( \mathcal{PT} \right)^{-1} =  e^{-i\phi},
\end{equation}
where $n_\sigma=0$ ($1$) for $\sigma=\uparrow$ ($\downarrow$)
and $\bar{\sigma}$ represents an opposite spin of $\sigma$.
By applying the $\mathcal{PT}$ operator on the Hamiltonian $H$,
we expand the equation as follows:
\begin{align}
	\left( \mathcal{PT} \right) H \left( \mathcal{PT} \right)^{-1}
	&= - t \sum_{l=0}^{L-2}\sum_{\sigma}
	\left( \mathcal{PT} \right) \left(
	e^{A} c^{\dagger}_{l+1,\sigma}  c_{l,\sigma} +
	e^{-A} c^{\dagger}_{l,\sigma}  c_{l+1,\sigma} 
	\right) \left( \mathcal{PT} \right)^{-1}
	+  U \sum_{l=0}^{L-1} \left( \mathcal{PT} \right)
	n_{l,\uparrow} n_{l,\downarrow} \left( \mathcal{PT} \right)^{-1}
	\nonumber \\
	& \quad -t \sum_\sigma
	\left( \mathcal{PT} \right) \left( 
	e^{A+i\phi} c^{\dagger}_{0,\sigma}  c_{L-1,\sigma} +
	e^{-A-i\phi} c^{\dagger}_{L-1,\sigma}  c_{0,\sigma} 
	\right) \left( \mathcal{PT} \right)^{-1}
	\nonumber \\
	&= - t \sum_{l=0}^{L-2}\sum_{\sigma}
	\left(
	e^{A} c^{\dagger}_{L-2-l,\bar{\sigma}}  c_{L-1-l,\bar{\sigma}} +
	e^{-A} c^{\dagger}_{L-1-l,\bar{\sigma}}  c_{L-2-l,\bar{\sigma}} 
	\right)
	+ U \sum_{l=0}^{L-1} n_{L-1-l,\downarrow} n_{L-1-l,\uparrow}
	\nonumber \\
	& \quad -t \sum_\sigma
	\left( 
	e^{A-i\phi} c^{\dagger}_{0,\bar{\sigma}}  c_{L-1,\bar{\sigma}} +
	e^{-A+i\phi} c^{\dagger}_{0,\bar{\sigma}}  c_{L-1,\bar{\sigma}} 
	\right).
\end{align}
By replacing the site index ($l' = L-2-l$ and $l'' = L-1-l$) and spin state 
($\sigma'=\bar{\sigma}$),
we get the equation as follows:
\begin{align}
	\left( \mathcal{PT} \right) H \left( \mathcal{PT} \right)^{-1} &=
	- t \sum_{l'=0}^{L-2}\sum_{\sigma'}
	\left(
	e^{A} c^{\dagger}_{l',\sigma'}  c_{l'+1,\sigma'} +
	e^{-A} c^{\dagger}_{l'+1,\sigma'}  c_{l',\sigma'} 
	\right)
	+ U \sum_{l''=0}^{L-1} n_{l'',\uparrow} n_{l'',\downarrow}
	\nonumber \\
	& \quad -t \sum_{\sigma'}
	\left( 
	e^{A-i\phi} c^{\dagger}_{L-1,\sigma'}  c_{0,\sigma'} +
	e^{-A+i\phi} c^{\dagger}_{0,\sigma'}  c_{L-1,\sigma'} 
	\right) \nonumber \\
	&= H^\dagger.
\end{align}
Thus, $H$ is the psuedo-Hermitian operator with respect to
the $\mathcal{PT}$ symmetric operator.
Consequently, the eigenvalues for the twisted (periodic) boundary condition
should either be real or come in complex-conjugate pairs~\cite{Mostafazadeh2002a}.

\subsection{Symmetry}

Let $\hat{N}$ and $\hat{S}_z$ be operators for the total number of fermions
and the $z$-component of the total spin,
given as $\hat{N}=\sum_{l=0}^{L-1}\left( n_{l,\uparrow}+n_{l,\downarrow} \right)$ and
$\hat{S}_z = \sum_{l=0}^{L-1} \left(n_{l,\uparrow}-n_{l,\downarrow}\right)$.
Importantly, the Hamiltonian $H$ commutes both $\hat{N}$ and $\hat{S}_z$ operators,
regardless of the selected $A$ value and boundary condition.
This implies that we can determine the eigenvalues of the Hamiltonian 
by considering the subspace 
in which both $\hat{N}$ and $\hat{S}_z$ are constant.
For $N$-fermion systems,
we can conveniently adopt basis states for the Hilbert space as follows:
\begin{equation}
	\left| \Psi_n \right> = \prod_{j=1}^N c^{\dagger}_{l_{nj},\sigma_{nj}}\left| vac \right>,
	\label{Eq_base}
\end{equation}
where $l_{nj}$ and $\sigma_{nj}$ denote the site index and spin state of 
the $j$th fermion for the $n$th state, respectively, 
and $\left| vac \right>$ represents the vacuum state.

In periodic and twisted boundary conditions, the Hamiltonian exhibits
the translation symmetry. We can take into account $L$ number of translation operators, 
defined as $T_{l}c^\dagger_{l',\sigma}T_{l}^{-1}=c^\dagger_{l'+l \mod L,\sigma}$ ($0 \le l,l' < L $).
$T_0$ corresponds to the identity operator.
With the translation symmetry, the many-body states can be
characterized with the crystal momentum $k \in \{0,2\pi/L,\cdots,2\pi(L-1)/L\}$.
The translation-symmetrized basis states can be expressed as follows:
\begin{equation}
	\left| \Psi_n \left(k\right) \right> = \frac{1}{\sqrt{N_{nk}}} \sum_{l=0}^{L-1}
	e^{-ikl} T_l \left| \Psi_n \right>,
\end{equation}
where $N_{nk}$ is the normalization factor and 
$\left| \Psi_n \right>$ is the representative state for the $n$th basis state.
$T_l \left| \Psi_n \right>$ can be obtained as follows:
\begin{equation}
	T_l \left| \Psi_n \right> = 
	\prod_{j=1}^N T_l c^{\dagger}_{l_{nj},\sigma_{nj}} T_l^{-1} \left| vac \right>
	= \prod_{j=1}^N c^{\dagger}_{l_{nj}+l \mod L,\sigma_{nj}} \left| vac \right>.
\end{equation}
By utilizing the translation symmetry and expressing the basis states 
in terms of crystal momentum, we can effectively reduce the Hilbert space, 
making it more tractable to calculate the eigenvalues and eigenstates 
of the Hamiltonian. 

\subsection{Non-interacting case (U=0)}

Let $H_t$ be the Hamiltonian for $U=0$ which is given as
\begin{equation}
	H_t = - t \sum_{l=0}^{L-2}\sum_{\sigma}
	\left( e^{A} c^{\dagger}_{l+1,\sigma}  c_{l,\sigma} +
	e^{-A} c^{\dagger}_{l,\sigma}  c_{l+1,\sigma} \right)
	+ H_{B}.
\end{equation}
To verify the normality of non-Hermitian $H_t$, 
we calculate the commutator $\left[ H_t , H_t^\dagger \right]$ as following:
\begin{align}
	\left[ H_t , H_t^\dagger  \right] &= t^2 \sum_{l=1}^{L-2}\sum_{\sigma}
	\left\{ 
	e^{2A} \left( c_{l+1,\sigma}^\dagger c_{l+1,\sigma} - c_{l,\sigma}^\dagger c_{l,\sigma} \right)
	+ e^{-2A}  \left( c_{l,\sigma}^\dagger c_{l,\sigma} - c_{l+1,\sigma}^\dagger c_{l+1,\sigma} \right)
	\right\}
	\nonumber \\
	&+ s t^2 \sum_\sigma
	\left\{ 
	e^{2A} \left( c_{0,\sigma}^\dagger c_{0,\sigma} - c_{L-1,\sigma}^\dagger 
	c_{L-1,\sigma} \right)
	+ e^{-2A}  \left( c_{L-1,\sigma}^\dagger c_{L-1,\sigma} - c_{0,\sigma}^\dagger c_{0,\sigma} \right)
	\right\} \nonumber \\
	&= 2 t^2 \sinh \left( 2A \right) \left( 1 - s\right) \sum_{\sigma}
	\left( c_{L-1,\sigma}^\dagger c_{L-1,\sigma} - c_{0,\sigma}^\dagger c_{0,\sigma} \right)
	,
\end{align}
where $s=1$ ($s=0$) for the periodic (open) boundary condition.
Notably, when $\left[ H_t , H_t^\dagger  \right]=0$ is satisfied,
$H_t$ becomes a normal operator, enabling us to construct
orthonormal right and left eigenstates, even if the eigenvalues may be complex.
Consequently, all eigenstates for the periodic boundary 
are orthonormal each other when $U=0$.
However, for the open boundary condition, 
$\left[ H_t , H_t^\dagger  \right]$ is no longer zero when $A \ne 0$.
In this case, the orthonormality of right and left eigenstates is not mandatory.

\subsection{Open boundary condition}

In open boundary condition, 
the one-dimensional Hatano-Nelson model of spin-half fermions is given as
\begin{equation}
	H = - t \sum_{l=0}^{L-2}\sum_{\sigma}
	\left(
	e^{A} c^{\dagger}_{l+1,\sigma}  c_{l,\sigma} +
	e^{-A} c^{\dagger}_{l,\sigma}  c_{l+1,\sigma} \right)
	+ U \sum_{l=0}^{L-1} c^\dagger_{l,\uparrow} c_{l,\uparrow} 
	c^\dagger_{l,\downarrow} c_{l,\downarrow}.
	\label{Eq_Hubbard}
\end{equation}
For the non-orthonomality property of eigenstates, we introduce new local fermionic operators $g^\dagger_{l,\sigma}$ and $\bar{g}_{l,\sigma}$,
defined as $g^\dagger_{l,\sigma} = e^{ lA} c^{\dagger}_{l,\sigma}$
and $\bar{g}_{l,\sigma} = e^{- l A} c_{l,\sigma}$.
The Hamiltonian of Eq.~\ref{Eq_Hubbard} is rewritten with the psuedo-Hermitian form as following:
\begin{equation}
	H=- t \sum_{l=0}^{L-2}\sum_{\sigma}
	\left(
	g^{\dagger}_{l+1,\sigma}  \bar{g}_{l,\sigma} +
	g^{\dagger}_{l,\sigma}  \bar{g}_{l+1,\sigma} \right) 
	+ U \sum_{l=0}^{L-1} g^\dagger_{l,\uparrow} \bar{g}_{l,\uparrow} 
	g^\dagger_{l,\downarrow} \bar{g}_{l,\downarrow}.
	\label{Eq_herm}
\end{equation}
In contrast, in periodic (twisted) boundary conditions, the boundary Hamiltonian $H_B$ is written as following:
\begin{equation}
	H_B = - t \sum_{\sigma} 
	\left( e^{L A} g^{\dagger}_{0,\sigma}  \bar{g}_{L-1,\sigma} +
	e^{-L A} g^{\dagger}_{L-1,\sigma}  \bar{g}_{0,\sigma} \right).
\end{equation}
Because of uncompensated phase terms $e^{\pm L A}$, $H_B$ cannot be expressed with the pseudo-Hermitian form like Eq.~\ref{Eq_herm}.

When the imaginary vector potential $A$ is set to be zero ($A=0$),
the Hamiltonian of Eq.~\ref{Eq_Hubbard} becomes Hermitian,
resulting in real eigenvalues and orthonormal eigenstates.
The eigenstates of the Hamiltonian $H_{A=0}$ (Hamiltonian with $A=0$) are
given as $H_{A=0} \left| \Psi_E^{0} \right> = E \left| \Psi_E^{0} \right>$,
and it can be expressed with the linear combination of orthonormal basis states:
$\left| \Psi_E^{0} \right> = \sum_{n} C^E_n \left| \Psi_n \right>$,
where $E$ is the eigenvalue and $\left| \Psi_n \right>$ is defined in Eq.~\ref{Eq_base}.
The coefficients $C_n^E$ can be determined by solving the following eigenproblem:
\begin{equation}
	\sum_{m} h^0_{nm} C^E_m = E  C^E_n,
	\label{Eq_eigen}
\end{equation}
where $h^0_{nm}= \left< \Psi_n \right| H_{A=0} \left| \Psi_m \right>$.
Because $H_{A=0}$ is the real-value operator, we can get $h^{0}_{nm}=h^{0}_{mn}$.

In the case when $A \ne 0$, the Hamiltonian is no longer Hermitian
due to non-zero non-reciprocal hopping, possibly leading to 
complex eigenvalues and non-orthonormal right and left eigenstates.
We introduce non-orthonormal basis states defined as follows:
\begin{subequations}
	\begin{equation}
		\left| \Psi^R_n \right> =
		\prod_{j=1}^N g^{\dagger}_{l_{nj},\sigma_{nj}}\left| vac \right>
		= e^{A \sum_{j=1}^N l_{nj}}\left| \Psi_n \right>,
	\end{equation}
	\begin{equation}
		\left| \Psi^L_n \right> =
		\prod_{j=1}^N \bar{g}^{\dagger}_{l_{nj},\sigma_{nj}}\left| vac \right>
		= e^{-A \sum_{j=1}^N l_{nj}}\left| \Psi_n \right>,
	\end{equation}
	\label{Eq_baseRL}
\end{subequations}
to express non-orthonormal right and left eigenstates of Eq.~\ref{Eq_Hubbard}, respectively.
Considering Eq.~\ref{Eq_base} and \ref{Eq_baseRL},
we can deduce that $\left< \Psi^L_n \right| g^\dagger_{l,\sigma} \bar{g}_{l',\sigma'} 
\left| \Psi^R_m \right> =
e^{A\left(l-l'\right)} e^{A\sum_{j=1}^N\left(l_{mj}-l_{nj} \right)}
\left< \Psi_n \right| 
c^{\dagger}_{l,\sigma}  c_{l',\sigma'}  \left| \Psi_m \right>$.
Here, the term
$\left< \Psi_n \right| c^{\dagger}_{l,\sigma}  c_{l',\sigma'}  \left| \Psi_m \right>$
has a non-zero value only when 
$l \in \{{ l_{n1},\cdots, l_{nN} \}}$, $l' \in \{{ l_{m1}, \cdots, l_{mN} \}}$,
and $\{{ l_{n1},\cdots, l_{nN} \}} - \{{ l \}} = \{{ l_{m1}, \cdots, l_{mN} \}} - \{{ l' \}}$.
We can get $\left< \Psi_n \right| 
c^{\dagger}_{l,\sigma}  c_{l',\sigma'}  \left| \Psi_m \right> =
\left< \Psi^L_n \right| g^\dagger_{l,\sigma}  \bar{g}_{l',\sigma'} 
\left| \Psi^R_m \right> $.
As a result, we can easily derive following relation:
\begin{equation}
	\left< \Psi_n \right| H_{A=0} \left| \Psi_m \right> =
	\left< \Psi_m \right| H_{A=0} \left| \Psi_n \right> =
	\left< \Psi^L_n \right| H \left| \Psi^R_m \right>
	= \left< \Psi^R_m \right| H^\dagger \left| \Psi^L_n \right>
	= \left< \Psi^R_n \right| H^\dagger \left| \Psi^L_m \right>,
	\label{Eq_HAZ}
\end{equation}
because $H_{A=0}$ is Hermitian.

Let $E_R$ ($E_L$) and $\left| \Psi^R_{E_R} \right>$ ($\left| \Psi^L_{E_L} \right>$) 
be the right (left) eigenvalue and eigenstate of Eq.~\ref{Eq_Hubbard}, respectively.
Utilizing the eigenvalue equations $H \left| \Psi^R_{E_R} \right> = E_R \left| \Psi^R_{E_R} \right>$
($ H^\dagger \left| \Psi^L_{E_L} \right>  = E_L^* \left| \Psi^L_{E_L} \right>$),
along with the biorthonormal property
$\sum_n \left| \Psi^R_n \right> \left< \Psi^L_n \right|
= \sum_n \left| \Psi_n \right> \left< \Psi_n \right| = \mathbf{I}$
($\sum_n \left| \Psi^L_n \right> \left< \Psi^R_n \right|
= \sum_n \left| \Psi_n \right> \left< \Psi_n \right| = \mathbf{I}$),
we obtain the following equations:
\begin{subequations}
	\begin{equation}
		\sum_{m} \left< \Psi^L_n \right| H  \left| \Psi^R_m \right>
		\left< \Psi^L_m \big| \Psi^R_{E_R} \right>
		= E_R \left< \Psi^L_n \big| \Psi^R_{E_R} \right>,
	\end{equation}
	\begin{equation}
		\sum_{m} \left< \Psi^R_n \right| H^\dagger  \left| \Psi^L_m \right>
		\left< \Psi^R_m \big| \Psi^L_{E_L} \right>
		= E_L^* \left< \Psi^R_n \big| \Psi^L_{E_L} \right>.
	\end{equation}
	\label{Eq_eigenA}
\end{subequations}
Since Eq.~\ref{Eq_HAZ} and \ref{Eq_eigenA},
we deduce the following eigenproblems:
\begin{subequations}
	\begin{equation}
		\sum_{m} h^0_{nm} \left< \Psi^L_m \big| \Psi^R_{E_R} \right>
		= E_R \left< \Psi^L_n \big| \Psi^R_{E_R} \right>,
	\end{equation}
	\begin{equation}
		\sum_{m} h^0_{nm} \left< \Psi^R_m \big| \Psi^L_{E_L} \right>
		= E_L^* \left< \Psi^R_n \big| \Psi^L_{E_L} \right>.
	\end{equation}
\end{subequations}
Equation~\ref{Eq_eigen} supports the solution with $E_R = E_L^* = E$ and
$\left< \Psi^L_n \big| \Psi^R_{E_R} \right> = 
\left< \Psi^R_n \big| \Psi^L_{E_L} \right> = C_n^E$.
The left and right eigenvalues of Eq.~\ref{Eq_Hubbard} in open boundary condition
are purely real and remain robust regardless of the strength of $A$.
Right and left eigenstates are given as following:
\begin{subequations}
	\begin{equation}
		\left| \Psi_E^R \right> = \sum_{n} C^E_n \left| \Psi_n^R \right>
		=\sum_{n} e^{A\sum_j^N l_{nj}} C^E_n \left| \Psi_n \right>,
	\end{equation}
	\begin{equation}
		\left| \Psi_E^L \right> = \sum_{n} C^E_n \left| \Psi_n^L \right>
		=\sum_{n} e^{-A\sum_j^N l_{nj}} C^E_n \left| \Psi_n \right>.
	\end{equation}
\end{subequations}
In contrast with the Hermitian case $(A=0)$,
eigenstates are not orthonormal each other when $A \ne 0$ 
due to the factor $e^{A\sum_j^N l_{nj}}$ ($e^{-A\sum_j^N l_{nj}}$) of 
basis state $\left| \Psi_n^R \right>$ ($\left| \Psi_n^L \right>$).

The complete set of real eigenvalues of non-Hermitian Hamiltonians is directly 
linked to the concept of pseudo-Hermiticity of Hamiltonian~\cite{Mostafazadeh2002b}.
To explore this, we introduce an invertible linear operator $S$ defined as:
\begin{equation}
	S = \sum_n \left| \Psi_n^R \right> \left< \Psi_n \right|,
	\label{Eq_S}
\end{equation}
where $\left| \Psi_n \right>$ and $\left| \Psi_n^R \right>$ are 
the $n$th orthonormal basis states and non-orthonormal basis states for right eigenstates of $H$, respectively.
Utilizing the biorthonormal property, we can verify 
that the inverse operator of $S$ is given as:
\begin{equation}
	S^{-1} = \sum_n \left| \Psi_n \right> \left< \Psi_n^L \right|,
	\label{Eq_SI}
\end{equation}
where $\left| \Psi_n^L \right>$ is the $n$th non-orthonormal basis states for left eigenstate of $H$.
Utilizing Eqs.~\ref{Eq_HAZ}, \ref{Eq_S} and \ref{Eq_SI}, 
we can deduce the following relation:
\begin{equation}
	\left< \Psi_n \right| S^{-1} H S \left| \Psi_m \right> =
	\sum_{n'm'}
	\left< \Psi_n \vert \Psi_{n'}\right>
	\left< \Psi_{n'}^L \right| H \left| \Psi_{m'}^R \right> 
	\left< \Psi_{m'} \vert \Psi_{n}\right>
	= \left< \Psi^L_n \right| H \left| \Psi^R_m \right>
	= \left< \Psi_n \right| H_{A=0} \left| \Psi_m \right>.
\end{equation}
We arrive at the following relation:
\begin{equation}
	H_{A=0} = S^{-1} H S,
\end{equation}
where $H_{A=0}$ is the Hermitian operator.
Moreover, we obtain $S S^\dagger = 
\sum_{n}\left| \Psi_n^R \right> \left< \Psi_n^R \right|$ and
$ \left( S S^\dagger \right)^{-1} = 
\sum_{n}\left| \Psi_n^L \right> \left< \Psi_n^L \right|$.
We can deduce the following relation:
\begin{equation}
	\left< \Psi^R_n \right| \left( S S^\dagger\right)^{-1}
	H S S^\dagger  \left| \Psi^L_m \right>
	= \sum_{n'm'} \left< \Psi^R_n \big| \Psi^L_{n'}\right>
	\left< \Psi^L_{n'} \right| H \left| \Psi^R_{m'} \right>
	\left< \Psi^R_{m'} \big| \Psi^L_{m}\right>
	=
	\left< \Psi^L_n \right| H \left| \Psi^R_m \right>
	= \left< \Psi^R_n \right| H^\dagger \left| \Psi^L_m \right>,
\end{equation}
where we utilize Eq.~\ref{Eq_HAZ}.
We can prove that $H$ is the $S S^{\dagger}$-pseudo-Hermitian
such that $H=\left( S S^{\dagger} \right) H^\dagger 
\left( S S^{\dagger} \right)^{-1}$.
According to A. Mostafazadeh's theorem~\cite{Mostafazadeh2002b},
we can verify that all eigenvalues of $H$ are the same as those of $H_{A=0}$
for the open boundary conditions, giving real values.
Furthermore, by utilizing Eq.~\ref{Eq_eigenA}, we can find the following relations:
\begin{subequations}
	\begin{equation}
		\left| \Psi_E^R \right> = \sum_{n} C^E_n \left| \Psi_n^R \right>
		=\sum_{n}  \big| \Psi_{n}^R \big> \big< \Psi_n \big| \Psi_E^0 \big>
		= S \big| \Psi_E^0 \big>,
	\end{equation}
	\begin{equation}
		\left< \Psi_E^L \right| = \sum_{n} (C^E_n)^* \big< \Psi_n^L \big|
		=\sum_{n} \big< \Psi_E^0 \big| \Psi_n \big> \big< \Psi_n^L \big|
		= \big< \Psi_E^0 \big| S^{-1}.
	\end{equation}
	\label{Eq_SPsi}
\end{subequations}
We conclude that right (left) eigenstates for finite $A$ can be obtained
using the similarity transformation from orthonormal eigenstates of $H_{A=0}$ with respect to the invertible linear operator $S$ ($S^{-1}$) under open boundary conditions.

\section{Non-Hermitian skin effect}

As proven in previous subsection, for open boundary conditions, the right eigenstates of $H$ can be always obtained 
by the similarity transformation from orthonormal eigenstates of $H_{A=0}$ with respect to the linear operator 
$S = \sum_{n} \big| \Psi_n^R \big> \big< \Psi_n \big| = 
\sum_{n} e^{A\sum_j^N l_{nj}} \big| \Psi_n \big> \big< \Psi_n \big|$.
For the reason, the coefficients of right eigenstates with respect to the $n$th orthonormal basis state $\left| \Psi_n \right>$ grow exponentially due to
the non-orthonormal factor  $e^{A\sum_j^N l_{nj}}$
when the magnitude of the imaginary vector potential $A$ increases.
In one-particle system, it leads to the exponential localization of all eigenstates at a specific boundary. This phenomena is commonly known as the non-Hermitian skin effect~\cite{Hatano1996,Hatano1998}.
In many-fermion systems, however, the localization of fermions is 
influenced not only by the exponential factor of eigenstates but also 
by the fermionic statistics governed by the Pauli exclusion principle.
This interplay gives rise to the many-body skin effect, 
which can be characterized by the asymmetry of
number distribution over open-boundary chain~\cite{Lee2020,Zhang2022,Alsallom2022,Kawabata2022}.
To examine this effect, we can calculate the number distribution functions
of right and left eigenstates using 
$n_E^R \left( l \right) = 
\left< \Psi_E^R \right|\sum_{\sigma} c^\dagger_{l,\sigma} c_{l,\sigma} \left| \Psi_E^R \right>/
\left< \Psi_E^R \big| \Psi_E^R \right> $ and
$n_E^L \left( l \right) = 
\left< \Psi_E^L \right|\sum_{\sigma} c^\dagger_{l,\sigma} c_{l,\sigma} \left| \Psi_E^L \right>/
\left< \Psi_E^L \big| \Psi_E^L \right>$, respectively.
The asymmetric distribution leads to an imbalance in the total
number of fermions located below $L/2$ ($0 \le l < L/2$) 
and above $L/2$ ($L/2 \le l < L$) sites, 
which can be quantified by the following formula:
\begin{subequations}
	\begin{equation}
		\mathcal{I}_E^R =  \sum_{L/2 \le l < L} n_E^R \left( l \right) -
		\sum_{0 \le l < L/2} n_E^R \left( l \right).
	\end{equation}
	\begin{equation}
		\mathcal{I}_E^L =  \sum_{L/2 \le l < L} n_E^L \left( l \right) -
		\sum_{0 \le l < L/2} n_E^L \left( l \right).
	\end{equation}
\end{subequations}
Non-zero values of $\mathcal{I}_E^R$ ($\mathcal{I}_E^L$) provides 
the characteristic feature of the many-body skin effect
as determined in right (left) eigenstates.

In the half-filled case ($N=L$), the non-orthonormal factor 
$e^{A\sum_j^N l_j}$ can be expressed as
\begin{equation}
	e^{A\sum_j^N l_j} =
	e^{ A \sum_{l=0}^{L-1} l}
	e^{A \sum_{l=0}^{L-1}l \left( \delta_{n_l,2}- \delta_{n_l,0}\right) }
	= e^{A\frac{L(L-1)}{2}}
	e^{A \sum_{l=0}^{L-1}l \left( \delta_{n_l,2}- \delta_{n_l,0}\right) },
\end{equation}
where $n_l$ refers to the number of fermions occupied at the $l$th site.
Here, $\delta_{n_l,2} = 1$ and $\delta_{n_l,0}=1$ indicate 
the presence of doublon and holon at the $l$th site, respectively,
and the total number of doublons ($N_d = \sum_{l=0}^{L-1} \delta_{n_l,2}$) 
and holons ($N_h =\sum_{l=0}^{L-1} \delta_{n_l,0}$) are equal.
For states with no doublon-holon pair ($N_d=N_h=0$), 
their non-orthonormal factor reaches its minimum with a constant value of $e^{A\frac{L(L-1)}{2}}$. 
In this case, the skin effect is not involved.
In contrast, states with finite doublons and holons exhibit 
a varying non-orthonormal factor 
depending on the distribution of doublons and holons.
The larger the segregation between doublons and holons, the greater the non-orthonormal factor becomes.
Consequently, the many-body skin effect in the half-filled case
is characterized by the segregation of doublons and holes.

\section{Effective Hamiltonian in strong coupling limit}

\begin{figure}[!b]
	\centering
	\includegraphics[width=0.6\columnwidth]{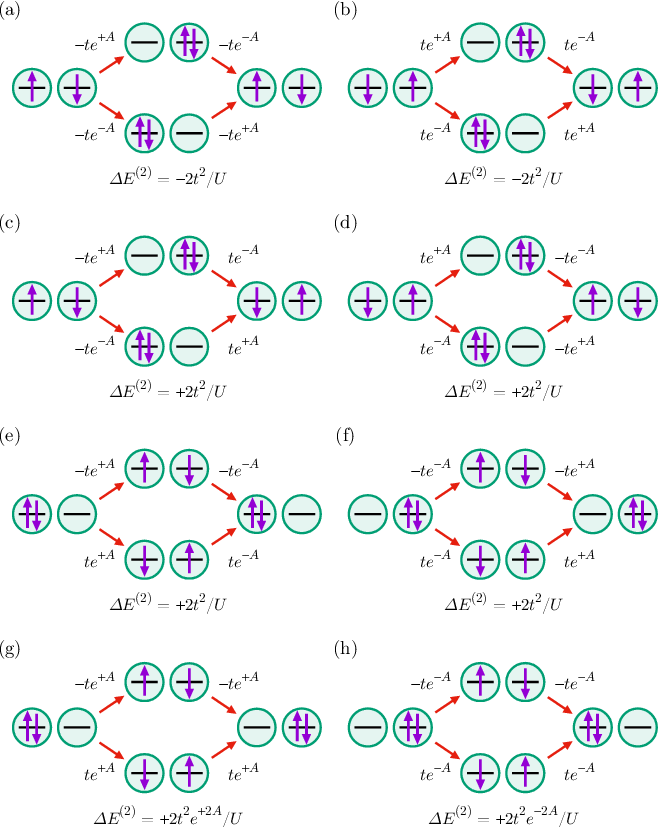}
	\caption {
		Schematic diagrams of relevant virtual hopping processes 
		to lead to the perturbation energy gain in the second-order perturbations.
		Diagrams in (a)--(d) present 
		the virtual hopping processes for the magnetic interactions between neighboring spins in the subspace $\mathcal{H}_0$,
		while diagrams in (e)--(f) describe the virtual hopping processes 
		for the perturbation energies between neighboring doublon-holon pairs in the subspace $\mathcal{H}_{L/2}$.
		$\Delta E^{(2)}$ below the diagram refers to the energy gain from 
		the virtual hopping process.
		Let the states for configurations depicted in the left, right, 
		and middle column of the diagrams be 
		$\left|\Psi_l \right>$, $\left| \Psi_r \right>$, and $\left| \Psi_m \right>$, respectively.
		$\Delta E^{(2)}= \pm \frac{1}{U} \sum_{m} 
		\left< \Psi_l \right| H_t \left| \Psi_m \right>  
		\left< \Psi_m \right| H_t \left| \Psi_r \right>$,
		where negative (positive) sign for $\mathcal{H}_{0}$ ($\mathcal{H}_{L/2}$).
	}
	\label{Sfig_Heff}
\end{figure}

In case of strong coupling limit ($U \gg t$), the dominant contribution to energy
arises from $H_U=U \sum_{l=0}^{L-1} n_{l,\uparrow}n_{l,\downarrow}$, while
the inclusion of $H_t = - t \sum_{l=0}^{L-1}\sum_{\sigma}
\left( e^{A} c^{\dagger}_{l+1,\sigma}  c_{l,\sigma} +
e^{-A} c^{\dagger}_{l,\sigma}  c_{l+1,\sigma} \right)$ leads to perturbation energy shifts.
Let $\mathcal{H}_{N_d}$ represent a subspace with a fixed number of doublons, $N_d$.
In the second-order perturbation theory, 
the Hamiltonian can be described using
the effective Hamiltonian within $\mathcal{H}_{N_d}$ as follows:
\begin{equation}
	H_{N_d}= N_d U + \mathcal{P}_{N_d} H_t \mathcal{P}_{N_d} + 
	\sum_{N'_d (N'_d \ne N_d)} \frac{\mathcal{P}_{N_d} H_t \mathcal{P}_{N'_d} H_t \mathcal{P}_{N_d}}{\left(N_d-N'_d\right)U},
\end{equation}
where $\mathcal{P}_{N_d}$ is a projection operator onto the subspace $\mathcal{H}_{N_d}{N_d}$~\cite{Essler2005}.
First, second, and third terms correspond to the center of energy of the subspace,
the kinetic energy contribution within the subspace, and the energy gain resulting from virtual hopping processes that connect different subspaces.
In the half-filled case, where the second term becomes zero,
we can derive the following equations:
\begin{subequations}
	\begin{equation}
		H_0 = - \frac{1}{U}\mathcal{P}_0 H_t \mathcal{P}_1 H_t \mathcal{P}_0,
	\end{equation}
	\begin{equation}
		H_{L/2} = \frac{L}{2}U + \frac{1}{U}\mathcal{P}_{L/2} H_t \mathcal{P}_{L/2-1} H_t \mathcal{P}_{L/2},
	\end{equation}
\end{subequations}
where $\mathcal{P}_0$, $\mathcal{P}_1$, $\mathcal{P}_{L/2-1}$, $\mathcal{P}_{L/2}$ are projection operators onto subspaces
$\mathcal{H}_0$ for no doublon-holon pair, $\mathcal{H}_1$ for one doublon-holon pair, $\mathcal{H}_{L/2-1}$ for $L/2-1$ doublon-holon pairs, and $\mathcal{H}_{L/2}$ for $L/2$ doublon-holon pairs, respectively.

As illustrated in Fig.~\ref{Sfig_Heff}(a)--(d), 
the perturbation energy gain is exclusively permitted within $\mathcal{H}_0$, 
only when neighboring fermions have opposite spin directions.
The effective Hamiltonian between fermions at $l$th and $(l+1)$th sites 
can be deduced as follows:
\begin{align}
	H_0^{l,l+1} &= - \frac{2t^2}{U} \big(
	\left| \uparrow_l,\downarrow_{l+1} \right> 
	\left< \uparrow_l,\downarrow_{l+1} \right| 
	+
	\left| \downarrow_l,\uparrow_{l+1} \right> 
	\left< \downarrow_l,\uparrow_{l+1} \right|
	-
	\left| \uparrow_l,\downarrow_{l+1} \right> 
	\left< \downarrow_l,\uparrow_{l+1} \right|
	-
	\left| \downarrow_l,\uparrow_{l+1} \right> 
	\left< \uparrow_l,\downarrow_{l+1} \right| 
	\big) \nonumber \\
	&= \frac{4t^2}{U} \left( \mathbf{S}_l \cdot \mathbf{S}_{l+1} - \frac{1}{4} \mathbf{I} \right).
\end{align}
Here, three spin operators at the $l$th site are given by:
$S_{l,x}=\frac{1}{2} \left(
\left| \uparrow_l \right> \left< \downarrow_l \right| +
\left| \downarrow_l \right> \left< \uparrow_l \right|
\right)$,
$S_{l,y}=\frac{i}{2} \left( 
-\left| \uparrow_l \right> \left< \downarrow_l \right| +
\left| \downarrow_l \right> \left< \uparrow_l \right|
\right)$, and
$S_{l,z}=\frac{1}{2} \left(
\left| \uparrow_l \right> \left< \uparrow_l \right| -
\left| \downarrow_l \right> \left< \downarrow_l \right|
\right)$.
In this effective Hamiltonian,
the non-reciprocal factor $e^{\pm A}$ cancels out
during the hopping process, where a fermion hops back and forth between neighboring sites.
This observation supports the stabilization of real eigenvalues 
within states predominantly characterized by the absence of doublon-holon pairs
in strong interaction limit.

To describe the effective Hamiltonian in the subspace $\mathcal{H}_{L/2}$, 
we introduce the creation (annihilation) operator of a doublon at the $l$th site
denoted as $b_{l}^\dagger = c_{l,\uparrow}^\dagger c_{l,\downarrow}^\dagger$
($b_{l} = c_{l,\downarrow} c_{l,\uparrow}$).
We can easily verify following commutation relations :
$\left[ b_{l}, b_{l'}^\dagger \right]=\delta_{l,l'}$, 
$\left[ b_{l}, b_{l'}\right]=\left[ b_{l}^\dagger, b_{l'}^\dagger \right]=0$,
and $b_{l}^\dagger b_{l} \left( 1-b_{l}^\dagger b_{l}\right) =0$.
As a result, the doublon can be treated as a hard-core boson.
Moreover, the perturbation energy gain is feasible
when doublon-holon pairs are located at neighboring sites 
as illustrated in Figs.~\ref{Sfig_Heff}(e)--(h). 
We can formulate the effective Hamiltonian between the $l$th and $(l+1)$th sites 
employing the hard-core boson representation:
\begin{align}
	H_{L/2}^{l,l+1} &= \frac{2t^2}{U} \left[
	b^\dagger_{l} b_{l} \left( 1- b^\dagger_{l+1} b_{l+1} \right) +
	b^\dagger_{l+1} b_{l+1} \left( 1- b^\dagger_{l} b_{l} \right) +
	e^{2A} b_{l+1}^\dagger b_{l} + e^{-2A} b_{l}^\dagger b_{l+1}
	\right] \nonumber \\
	&= \frac{2t^2}{U} \left( 
	e^{2A} b_{l+1}^\dagger b_{l} + e^{-2A} b_{l}^\dagger b_{l+1}
	\right) 
	- \frac{4t^2}{U} b^\dagger_{l} b_{l} b^\dagger_{l+1} b_{l+1}
	+ \frac{2t^2}{U} \left(
	b^\dagger_{l} b_{l}  + b^\dagger_{l+1} b_{l+1} \right).
\end{align}
Thus, we arrive at the effective Hamiltonians within 
$\mathcal{H}_0$ and $\mathcal{H}_{L/2}$ as follows:
\begin{subequations}
	\begin{equation}
		H_0 = J \sum_{l} \mathbf{S}_l \cdot \mathbf{S}_{l+1} - \frac{J}{4} L,
	\end{equation}
	\begin{equation}
		H_{L/2} = T_{\rm eff} \sum_l \left( 
		e^{A_{\rm eff}} b_{l+1}^\dagger b_{l} + 
		e^{-A_{\rm eff}} b_{l}^\dagger b_{l+1}
		\right)
		+ U_{\rm eff} \sum_{l} b^\dagger_{l} b_{l} b^\dagger_{l+1} b_{l+1}
		+\frac{L}{2} \left( U - U_{\rm eff} \right),
	\end{equation}
\end{subequations}
where $J=\frac{4t^2}{U}$, $T_{\rm eff} = \frac{2t^2}{U}$, $A_{\rm eff} =2A$, and
$U_{\rm eff} = - \frac{4t^2}{U}$, respectively.
Hence, $H_{L/2}$ can be regarded as the Hamiltonian of
half-filled hard-core bosons.
Here, adjacent hard-core bosons are attractively interacted by $U_{\rm eff} = - \frac{4t^2}{U}<0$
and their non-reciprocal hopping is given by the effective hopping strength $T_{\rm eff}= \frac{2t^2}{U}$ and 
the effective imaginary vector potential $A_{\rm eff}=2A$.

\section{Eigenspectrum at $U=0$}

Figures~\ref{Sfig_skin} and~\ref{Sfig_energy} depict the asymmetric
number density and eigenspectra of the non-interacting case, respectively.

\begin{figure}[h]
	\centering
	\includegraphics[width=0.7\columnwidth]{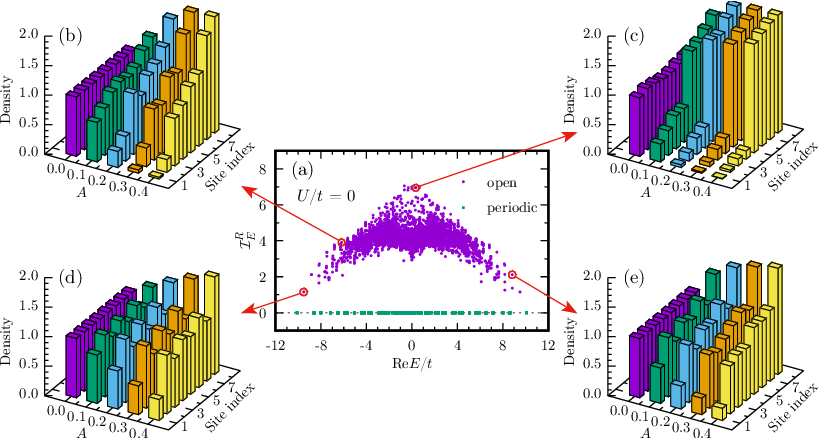}
	\caption {
		(a) The number density imbalance in the half-filled Hatano-Nelson model of spin-half fermions ($A=0.3$) for both open and periodic boundary conditions when $U/t=0$.
		The imbalance $\mathcal{I}_E^R$ is calculated by $\sum_{L/2 \le l < L} 
		n_E^R \left( l \right) - \sum_{0 \le l < L/2} n_E^R \left( l \right)$, where
		$n_E^R \left( l \right)$ is the number density of right eigenvalues at the $l$th site.
		(b)--(e) the number density distributions as a function of imaginary gauge potential $A$
		for specific right eigenstates indicated by red arrows
		pairs play a dominant role, respectively, for the open boundary condition.
		All results are obtained for the lattice size of eight ($L = 8$).
	}
	\label{Sfig_skin}
\end{figure}

\begin{figure}[h]
	\centering
	\includegraphics[width=0.5\columnwidth]{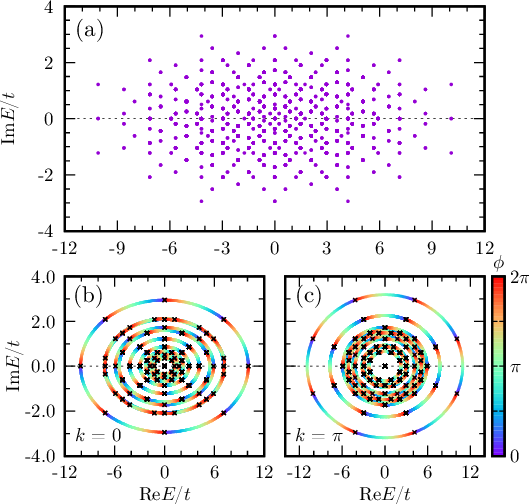}
	\caption {
		(a) Spectral distribution of eigenvalues in the half-filled Hatano-Nelson model of spin-half fermions when $U/t=0$ and $A=0.3$.
		Spectral change of eigenvalues as functions of the gauge flux $\phi$
		for the twisted boundary condition in (b) $k=0$, $S_z=0$ and (c) $k=\pi$, $S_z=0$. Here, $k$ is the crystal momentum of one dimensional lattice and $S_z$ is the $z$-component of the total spin.
		Cross points refer to the eigenvalues for $\phi=0$.
		Spectral data well depict the point-gap topology of complex eigenvalues.
		Point gap is located at the origin of complex plane.
		All results are obtained for the lattice size of eight ($L = 8$).
	}
	\label{Sfig_energy}
\end{figure}

\section{Orthonormlity property at the topological transitions}

We investigate the orthonormality properties of the eigenpairs 
involved in topological transitions by calculating 
the overlap between two right eigenstates of the eigenpair as follows:
\begin{equation}
	O_{nm}^R = 
	\frac{\left< \Psi^R_n \vert \Psi^R_m \right>\left< \Psi^R_m \vert \Psi^R_n \right>}
	{\left< \Psi^R_n \vert \Psi^R_n \right>\left< \Psi^R_m \vert \Psi^R_m \right>},
\end{equation}
where $n$ and $m$ are indices of right eigenvalues of the eigenpair.

In the case of $U=0$, the Hamiltonian is normal, 
resulting in all eigenstates being mutually orthonormal. 
For the reason, the overlap of all eigenpairs becomes to zero 
as shown in Fig.~\ref{Sfig_overU}.
However, when $U$ has finite values, this orthonormality no longer holds.
The overlap takes on finite values, and eventually reaches one
at the topological transition point.
This point is characterized by the merging of complex conjugate eigenvalues onto the real axis. 
Thus, two eigenstates of the eigenpair collapse into 
a singular state at the transition point, which are indeed the exceptional points.
Following the transition point, the overlap progressively decreases.
As $U$ approaches infinity, the overlaps converge into zero
because the Hermiticity is restored.

\begin{figure}[h]
	\centering
	\includegraphics[width=0.6\columnwidth]{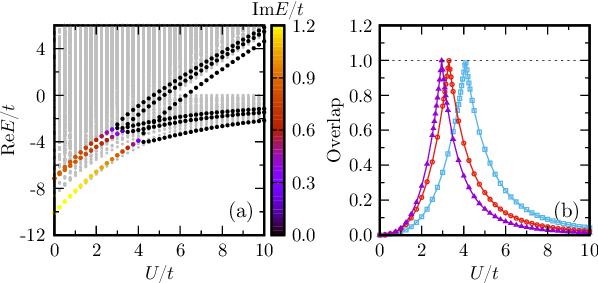}
	\caption {
		(a) Real components of eigenvalues as a function of Coulomb repulsion $U$
		for $0\le U/t \le 10$.
		Three examples of the topological transitions are highlighted by the color map.
		(b) Overlap between two eigenstates involved in the topological transitions.
		All results are obtained when $A=0.3$ and $L=8$.
	}
	\label{Sfig_overU}
\end{figure}

\section{Topological transitions under an imaginary gauge flux}

In the absence of an imaginary gauge flux ($A=0$), the Hamiltonian retains its Hermitianity.
This leads to purely real eigenvalues and full orthonormality of eigenstates.
When the magnitude of $A$ is large enough, however,
eigenvalues are able to form complex conjugate pairs, and the orthonormality among eigenstates is broken.
Therefore, we can conjecture that 
the topological transitions among eigenpairs, 
which are characterized by the existence of exceptional points,
take place by varying the magnitude of $A$.

Figure~\ref{Sfig_overA} presents well these topological transitions
in Mott insulating systems with $U/t=10$.
It is reminiscence of the breakdown of the Mott insulator under an imaginary gauge flux~\cite{Fukui1998}.
At this transition point, two real eigenvalues coalesce:
one in the zero-doublon-holon sector and the other in the one-doublon-holon sector.
Subsequently, eigenvalues branch out into complex conjugate pairs.
Because the transition occurs at the exceptional point,
the overlap between the eigenpair becomes one at the transition point.
These observed behaviors collectively support that
the $\mathcal{PT}$-like transitions, mediated by the imaginary gauge flux $A$,
occur through the exceptional points.

\begin{figure}[h]
	\centering
	\includegraphics[width=0.6\columnwidth]{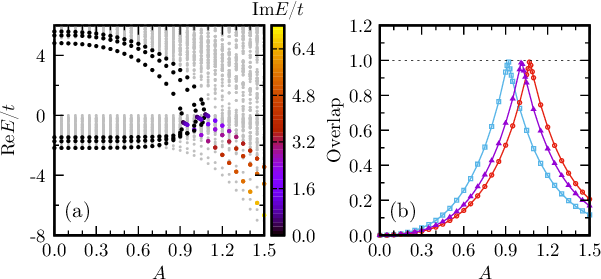}
	\caption {
		(a) Real components of eigenvalues as a function of imaginary gauge flux $A$
		for $0\le A \le 1.5$.
		Three examples of the topological transitions are highlighted by the color map.
		(b) Overlap between two eigenstates involved in the topological transitions.
		All results are obtained when $U/t=10$ and $L=8$.
	}
	\label{Sfig_overA}
\end{figure}

\end{widetext}
\end{document}